\documentclass[10.5pt,a4paper,oneside]{extarticle}
\usepackage[T1]{fontenc} 
\usepackage[left=25mm,right=20.9mm,top=4.25cm,bottom=3.75cm,headsep=0.25cm,footskip=0.25cm]{geometry}

\usepackage[english]{babel}
    \usepackage[affil-it]{authblk}
    \usepackage{etoolbox}
    \usepackage{lmodern}
    \usepackage{setspace}
    \usepackage{amssymb,amsthm,amsmath}
    \usepackage{tikz}
    \usepackage{listings}
    \usepackage{color}
    \usepackage{caption}
    \usepackage{placeins}
    \usepackage{booktabs}
    \usepackage{enumerate}
    \usepackage{graphicx}
    \usepackage{titlesec}
    \usepackage{mathptmx}
    \usepackage{anyfontsize}
    \usepackage{t1enc}
    \usepackage[utf8]{inputenc}
    \usepackage{fancyhdr}
    \usepackage{float,xspace}
    \usepackage{scalefnt}
    \usepackage{graphicx}
    \usepackage{subfigure}
    \usepackage{multirow}
\usepackage[round,sort,comma,numbers]{natbib}
\usepackage{soul} 
\usepackage{tabulary}
\usepackage[colorlinks=true,linkcolor=blue]{hyperref} 

\usepackage{textcase}

\usepackage[colorinlistoftodos]{todonotes}

\titlespacing*{\section}
{0pt}{12pt}{0pt}
\titlespacing*{\subsection}
{0pt}{12pt}{0pt}
\titlespacing*{\subsubsection}
{0pt}{12pt}{0pt}
\titleformat{\section}
  {\huge\sffamily\Large\bfseries\color{black}}
  {\thesection}{1em}{}
    \titleformat{\subsection}[block]{\bfseries}{\thesubsection}{.5em}{}
    \titleformat{\subsubsection}[block]{\bfseries}{\thesubsubsection}{.5em}{}

\titleformat{\section}{\fontsize{12}{19}\bfseries}{\thesection}{1em}{}

\usepackage{mathptmx} 

\makeatletter

\patchcmd{\@maketitle}{\LARGE \@title}{\fontsize{14}{19.2}\selectfont\@title}{}{} 
\makeatother
\title
{
	\vspace{-5cm}
	\begin{minipage}{\textwidth}	
	\end{minipage}
	\\[0.5cm] 
	\textbf{Domestic sound event detection by shift consistency mean-teacher training and adversarial domain adaptation}
	\author[ ]{Fang-Ching Chen, Kuan-Dar Chen, Yi-Wen Liu}
  	\affil[]{National Tsing Hua University, Taiwan \{fumchin@gmail.com, Jerrychen445@gmail.com, ywliu@ee.nthu.edu.tw\}}
}
\date{}

\begin{document}

\clearpage
\setcounter{page}{1}
\maketitle
\thispagestyle{empty}

\subsection*{\fontsize{10.5}{19.2}\uppercase{\textbf{Abstract}}}
{\fontsize{10.5}{60}\selectfont 
Semi-supervised learning and domain adaptation techniques have drawn increasing attention in the field of domestic sound event detection
thanks to the availability of large amounts of unlabeled data and the relative ease to generate synthetic strongly-labeled data. In a previous work, several semi-supervised learning strategies were designed to boost the performance of a mean-teacher model. Namely, these strategies include shift consistency training (SCT), interpolation consistency training (ICT), and pseudo-labeling. However, adversarial domain adaptation (ADA) did not seem to improve the event detection accuracy further when we attempt to compensate for the domain gap between synthetic and real data.  In this research, we empirically found that ICT tends to pull apart the distributions of synthetic and real data in t-SNE plots. Therefore, ICT is abandoned while SCT, in contrast, is applied to train both the student and the teacher models. With these modifications, the system successfully integrates with an ADA network, and we achieve 47.2\% in the F1 score on the DCASE 2020 task 4 dataset, which is 2.1\% higher than what was reported in the previous work.
}

\noindent{\\ \fontsize{11}{60}\selectfont Keywords: Sound event detection, adversarial domain adaptation, semi-supervised learning}

\fontdimen2\font=4pt

\section{\uppercase{Introduction}}
In sound event detection (SED), it is desired for a model to determine the onset and offset of the events precisely. However, marking the time stamps of events in an audio clip is a time-consuming and labor-intensive task when anyone attempts to provide such \emph{strongly-labeled} data to train a sound event classifier. 
Due to the growing attention to SED, 
a competition called Detection and Classification of Acoustic Scenes and Events (DCASE) has been held annually since 2017.
Currently, there are two urgent issues to address in this field; the first is how to utilize unlabeled data effectively, and the second is how to utilize a large amount of strong-labeled synthetic data to real-world environment. For the first problem, several semi-supervised learning strategies (SSL) have been utilized in the past, including pseudo-labeling \cite{pseudo}, mean-teaching \cite{MT}, interpolation consistency training (ICT) \cite{ICT}, and Mixmatch \cite{Mixmatch}; these strategies have been proven to help the SED system learn from the unlabeled data efficiently. Beside semi-supervised training strategies, the architecture of the backbone of the SED model is also worth investigating. For neural network-based backbones, convolutional neural network (CNN) \cite{CNN}, convolutional recurrent neural network (CRNN) \cite{CRNN}, residual convolutional recurrent neural network (RCRNN) \cite{RCRNN} are commonly used.

To tackle the second problem (scarcity of strongly-labeled data), one may contemplate using synthetic 
mixture of clips of known sound events with background audio recordings.
Compared with unlabeled and weakly-labeled real data, such synthetic data are appealing in that the accurate time stamps of sound events can be automatically created. However, 
the synthetic mixtures may 
have acoustic mismatch to the real-world recordings. Such mismatch needs to be regarded as domain differences, and various domain adaptation techniques 
 \cite{Acrnn,two_stage,MMTDA} have been proposed to align the distributions of synthetic and real audio samples in the feature space and alleviate the domain-gap problem.

In this research, we studied an existing mean teacher-based SED model \cite{previous} (from our research group) and found that its semi-supervised learning strategies somehow conflict with adversarial domain adaptation (ADA). Some performance analysis and visualization of data distributions in the feature space led us to believe that ICT should be abandoned while SCT could be emphasized by deploying it to both teacher and student models. We call this new arrangement \emph{shift consistency mean-teacher} (SCMT) training. Thereby, we successfully integrated SCMT strategies with ADA and improved the SED performance in the DCASE 2020 task 4 dataset. 
The rest of this paper is organized as follows: in Sec.~2, we give an overview of the baseline model and our newly proposed methods. Then we introduce the dataset used in the experiments. In Sec.~3, the parameters of data pre-processing and experiment setup will be described. Finally, in Sec.~4, we discuss and compare the performance of different models and strategies, also outlining the mutual incompatibility between ICT and ADA with both features visualization and quantitative analysis.

\section{\uppercase{Methods}}
The existing mean-teacher model is hereafter referred to as \emph{our previous work} \cite{previous} and briefly reviewed in section 2.1. In sections 2.2 and 2.3, the changes we made in this research will be described.
\subsection{Baseline} 
\begin{figure}[!h]
\centering
\includegraphics[width=12cm]{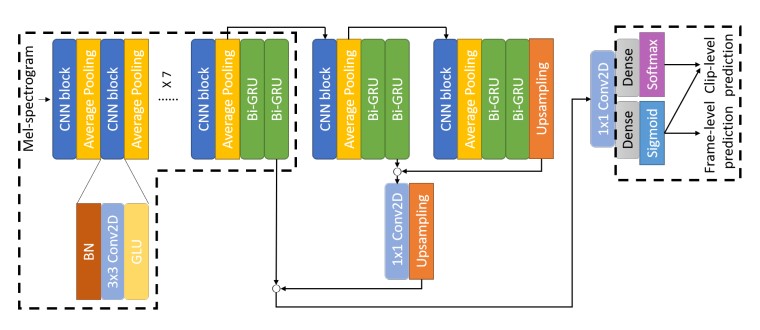}
\caption{Network architecture of FP-CRNN. The part enclosed by the dashed line is a CRNN, and the remaining parts are feature-pyramid components.}
\label{fig:figure1}
\end{figure}

In our previous work, a novel backbone and several effective mean-teacher based semi-supervised strategies were proposed and integrated. A convolutional recurrent neural network which contained feature-pyramid components (FP-CRNN) was used as the backbone; it was comprised of CNN blocks, bidirectional gated recurrent (GRU) \cite{GRU} unit cells, an attention part to generate the clip-level and frame-level output, and most importantly, feature-pyramid \cite{FP} components which were originally used in object detection task in computer vision. With the help of the feature-pyramid components, the model could utilize multi-scale features. Figure \ref{fig:figure1} shows the architecture of FP-CRNN.

The semi-supervised learning strategies that were utilized in our previous work included the mean-teacher technique, weakly pseudo-labeling, ICT, and SCT. Generally speaking, mean-teacher approaches contain a teacher model and a student model that share the same network architecture but have different weights. For the student model, the weights are learned from back propagation, while the weights of the teacher model are updated by the exponential moving average of the parameters of the student model. The loss function is given as follows, 
\begin{equation}
    L = L_{w,\text{BCE}} + L_{s,\text{BCE}} + r(t)(L'_{w,\text{MSE}}+L'_{s,\text{MSE}}),
    \label{eq:MT}
\end{equation}
where $t$ denotes the current step of training, 
\begin{equation}
    r(t) = \text{exp}\left[-5\left(1-\frac{t}{T}\right)^2\right]
\end{equation}
is a ramp-up function, and $T$ denotes the ramp-up length; the subscripts $w$ and $s$ denote clip-level output and frame-level output, respectively, $L$, $L'$ denote the loss between outputs of the student model and the ground truth and the loss between the outputs of the teacher model and the student model, respectively. Finally, MSE and BCE stands for mean squared error and binary cross entropy.

Next, in our previous work, a weakly pseudo-labeling (PL) technique was adopted to infer unlabeled data into weakly-labeled data by an audio tagging system which was pre-trained with weakly-labeled and strongly-labeled data. Generally speaking, compared with the accuracy in frame-level sound event detection, the accuracy in clip-level sound event classification is easier to improve by adopting a deeper neural network \cite{previous}. 
By generating reliable weak pseudo-labels, the performance of the sound event detection system could also be improved. Another semi-supervised learning strategy that was utilized is interpolation consistency training (ICT). It encouraged the prediction of the interpolation of unlabeled data to be consistent with the interpolation of predictions of the unlabeled data in the following manner; we desired that
\begin{equation}
    F_\theta(\text{Mix}_\lambda(u_j, u_k)) \approx \text{Mix}_{\lambda}(F_{\theta'}(u_j), F_{\theta'}(u_k)),
    \label{eq:ICT}
\end{equation}
where
 $   \text{Mix}_\lambda(a,b) = \lambda a + (1-\lambda) b$,
$u_j, u_k$ denote unlabeled data, $F_\theta$ denotes a student model and $F_{\theta'}$ denotes a teacher model. In our previous work, $\lambda$ was randomly sampled from Beta distribution. 

Though ICT helps the model to be consistent when facing ambiguous data, we still remove the ICT in this paper, and the reason will be given in section 4.3. Last but not the least, our previous work proposed \emph{shift consistency training} (SCT), which was inspired by ICT. It encouraged the predictions of time-shifted and pitch-shifted data to be consistent with time-shifted and pitch-shifted predictions. The loss function was designed as follows,
\begin{equation}
    L_{\rm{SCT}} = L_{wf,\rm{BCE}} +  L_{sf,\rm{BCE}} + L_{st,\rm{BCE}} + r(t)L_{st,\rm{MSE}},
    \label{eq:SCT}
\end{equation}
where subscripts $w$ and $t$ denote clip-level outputs (weak) and frame-level outputs (strong), respectively, and $f$ and $t$ denote frequency shift and time shift, respectively. 

\subsection{Shift consistency mean-teacher training}
\begin{figure}[h]
\centering
\includegraphics[width=16cm]{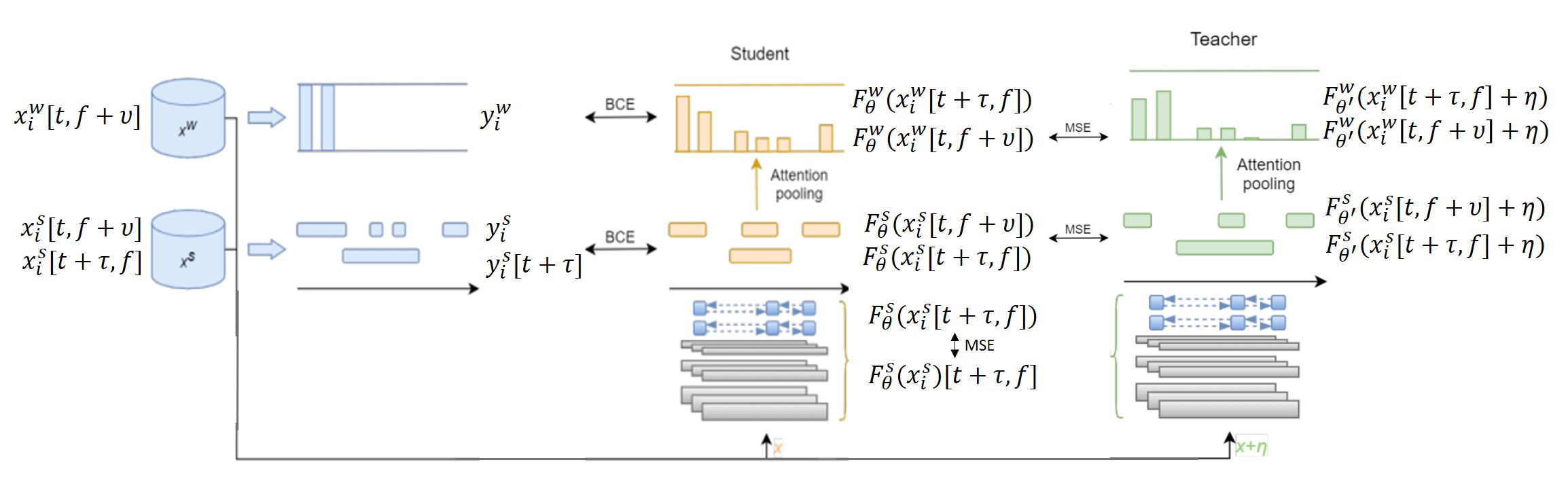}
\caption{Network architecture of SCMT. $\eta \sim \mathcal{N}(0, 0.5)$ denotes the Gaussian noise added to the input of the teacher model. $t$ and $f$ denote time and frequency, respectively; $\tau$ and $\nu$ denote the time and frequency shift, respectively. $F_\theta$ denotes a student model, $F_{\theta'}$ denotes a teacher model and the superscripts $w$ and $s$ denote clip-level and frame-level outputs from the model, respectively. }
\label{fig:figure2}
\end{figure}

SCT successfully helped the previous model to be more consistent by randomly shifting the spectrograms in time and frequency; however, it was only applied to the student model during training. In this paper, we propose a new version of SCT called shift consistency mean-teacher training (SCMT) by strengthening the relationship between the mean-teacher approach and SCT. For the implementation of the SCMT, we deploy SCT to both the student model and the teacher model. Figure \ref{fig:figure2} shows the concept and architecture of SCMT, and the loss function is devised as below,

\begin{equation}
    L_{\rm{SCMT}} = L_\text{SCT} + r(t)(L'_{wt,\rm{MSE}} + L'_{wf,\rm{MSE}} + L'_{st,\rm{MSE}} + L'_{sf,\rm{MSE}})
\end{equation}
where the subscript $w, s, f$ and $t$ have the same meanings as in Eq.~(\ref{eq:SCT}). 
\subsection{Adversarial domain adaptation}
Adversarial domain adaptation (ADA) \cite{DANN} refers to the broad idea to generate domain-invariant embedded features by fooling a domain discriminator $D$ with a generator $G_f$. During its training, a gradient reversal layer (GRL) \cite{GRL} acts as an identity transform during forward propagation, but changes the sign of the gradient before passing to the preceding layers during back-propagation. By placing GRL between $G_f$ and $D$, $G_f$ and $D$ are optimized in a reverse direction, while the label predictor $G_y$ will also be trained. The following shows the learning objective functions: 

\begin{equation}
    E(\theta_f, \theta_y, \theta_d) = L_{\rm{s}}(\theta_f, \theta_y) + L_{\rm{w}}(\theta_f, \theta_y) - \lambda_{\rm{d}} \cdot L_{\rm{d}, \rm{BCE}}(\theta_f, \theta_y, \theta_d),
\end{equation}
where ${\theta}_{f}$, ${\theta}_{y}$ and ${\theta}_{d}$ denote the parameters of $G_f$, $G_y$ and $D$, respectively. $\lambda_{\rm{d}}$ denotes the adversarial loss weight of the domain loss. $L_d$ denotes the frame-level domain loss. The subscript $w$, $s$ have the same meanings as in Eq.~(\ref{eq:MT}). The following equations describe the adversarial training process,
\begin{equation}
    (\hat{\theta}_{f}, \hat{\theta}_{y}) = \mathop{\arg\min}_{\theta_{f}, \theta_{y}}E(\theta_f, \theta_y, \hat{\theta}_d),
\end{equation}
and
\begin{equation}
    \hat{\theta}_{d} = \mathop{\arg\max}_{\theta_{d}}E(\hat{\theta}_f, \hat{\theta}_y, \theta_d).
\end{equation}

In the present research, we adopt the concept of two-stage \cite{two_stage} domain adaptation. In the first stage, we train the model that contains FP-CRNN and the semi-supervised learning strategies mentioned above to ensure the ability to detect the sound event classes. In the second stage, we add adversarial domain adaptation into the training to bridge the domain gap between real and synthetic data.

\section{\uppercase{Experiments}}
DCASE 2020 domestic environment sound event detection (DESED) dataset is used in the experiments. It provides ten different domestic sound event classes. The dataset consists of 10-second audio clips either recorded in domestic environments or in synthesized soundscapes. Each audio clip contains at least one sound event. For the training set, 1,578 weakly-labeled, 14,412 unlabeled real soundscapes, and 2,584 strongly-labeled synthetic soundscapes are provided. We evaluated the performance on the validation set which has 1,168 strongly-labeled real recordings.

Log-mel spectrograms were adopted as the input to the present model. To generate the spectrograms, all audio clips were first resampled to 16 kHz, while setting window size, hop length, and the number of mel bins to 2048, 255 and 128, respectively. Finally, for each audio clip, a $648 \times 128$ mel-log spectorgram was generated.

The amount of time shift and frequency shift for SCT and SCMT was normally distributed between $-2$ to $2$ sec and $-4$ to $4$ mel bins, respectively. As for weakly pseudo-labeling, we fine-tuned a pre-trained ResNet18 and used it to produce weak labels from unlabeled data. 

\section{\uppercase{Results and Discussion}}
All experiments adopted the pseudo-labeling strategy and the mean-teacher approach during training. 

\subsection{Evaluation of shift consistency mean-teacher model and domain adaptation}
Table \ref{tab:table1} shows that adversarial domain adaptation consistently improved the F1-score of all the models. Meanwhile, FP-CRNN with SCMT outperformed the FP-CRNN with SCT from our previous work by 2.0\% before ADA, and 3.5\% after ADA. Not only does this show that SCMT performs better than SCT, it also indicates that the compatibility between SCMT and ADA is better than between SCT and ADA. Also, Table \ref{tab:table2} shows the model we propose in this paper performed 2.1\% better than the highest F1-score that was reported in our previous work.

Table \ref{tab:table3}, however, shows that the performance of the models dropped when integrating both ICT and SCT with ADA. Thus, mutual compatibility between SSL strategies and ADA is a problem which needs to be further investigated. To this end, we applied t-SNE and silhouette analysis to visualize and quantify what happened in the embedded feature space when various combinations of learning techniques were applied.

\begin{table}[!h]
\caption{F1 score(\%) comparison of different models before and after domain adaptation}
\label{tab:table1}
\begin{center}
\begin{tabular}{ccccccc}
\hline 
Model & Before ADA & After ADA  \\ \hline \hline
CRNN & 38.2 & 40.3  \\
FP-CRNN & 40.8 & 42.1  \\ \hline
FP-CRNN + SCT & 42.7 & 43.7  \\
FP-CRNN + SCMT & 44.7 & \textbf{47.2}   \\ \hline
\end{tabular} 
\end{center}
\end{table}

\begin{table}[!h]
\caption{Comparison with the best-performing models from our previous work}
\label{tab:table2}
\begin{center}
\begin{tabular}{ccccccc}
\hline 
\multirow{2}{*}{Architecture} & \multicolumn{2}{c}{SSL Strategies}  &   \multirow{2}{*}{ADA}   & \multirow{2}{*}{F1 (\%)} \\ \cline{2-3} 
                              & ICT & SCT / SCMT  &  &   \\ \hline \hline
CRNN    &   v   &   SCT &   -   & \textbf{45.1} \\
FP-CRNN &   v   &   SCT &   -   & 44.5  \\ \hline
FP-CRNN &   -   &   SCMT&   v   & \textbf{47.2}  \\ \hline
\end{tabular} 
\end{center}
\end{table}

\begin{table}[!h]
\caption{Incompatibility between ADA and semi-supervised techniques. The F1 score(\%) of several architecture and strategies before and after domain adaptation is listed.}
\label{tab:table3}
\begin{center}
\begin{tabular}{ccccccc}
\hline 
Architecture & SSL Strategies & ADA  & F1 (\%)  \\ \hline \hline
CRNN    &   ICT+SCT  &   -   &   \textbf{45.1} \\
CRNN    &   ICT+SCT  &   v   &   \textbf{44.2}  \\ \hline
FP-CRNN &   ICT+SCT  &   -   &   \textbf{44.7} \\
FP-CRNN &   ICT+SCT  &   v   &   \textbf{43.0}  \\ \hline

\end{tabular} 
\end{center}
\end{table}

\subsection{Discussion on domain adaptation and its compatibility to semi-supervised strategies}
\begin{figure}[H]
\centering
\subfigure[FP-CRNN before domain adaptation]{
    \includegraphics[width=8cm]{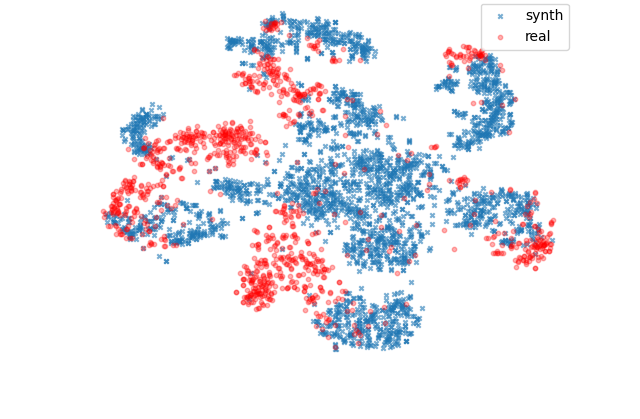}
    
}\subfigure[FP-CRNN after domain adaptation]{
    \includegraphics[width=7cm]{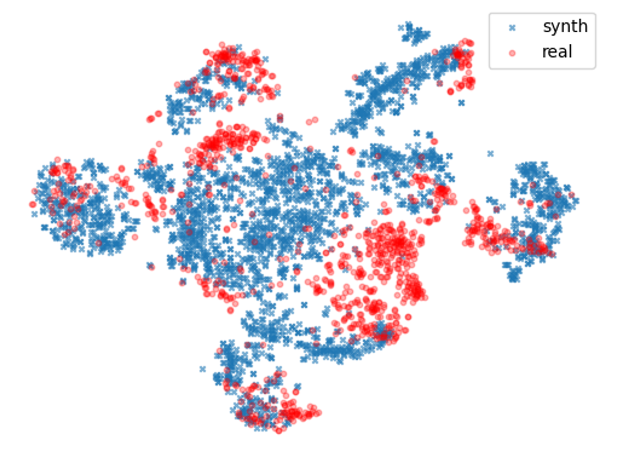}
}
\caption{t-SNE visualization of embedded features}
\label{fig:figure3}
\end{figure}

\begin{figure}[H]
\centering
\subfigure[FP-CRNN]{
    \includegraphics[width=5cm]{visualize/fpn.png}
}\subfigure[FP-CRNN + ICT]{
    \includegraphics[width=5cm]{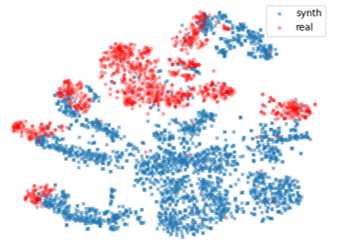}
}\subfigure[FP-CRNN + ICT + ADA]{
    \includegraphics[width=5cm]{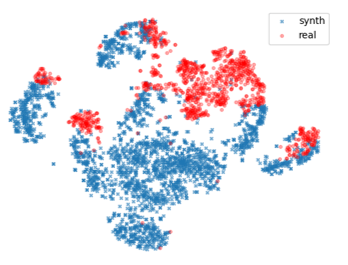}
}
\caption{t-SNE visualization of embedded features (ICT)}
\label{fig:figure4}
\end{figure}

\begin{figure}[H]
\centering
\subfigure[FP-CRNN]{
    \includegraphics[width=5cm]{visualize/fpn.png}
}\subfigure[FP-CRNN + SCT]{
    \includegraphics[width=5cm]{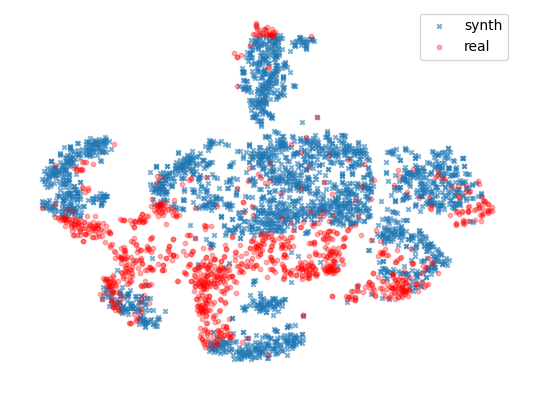}
}\subfigure[FP-CRNN + SCT + ADA]{
    \includegraphics[width=5cm]{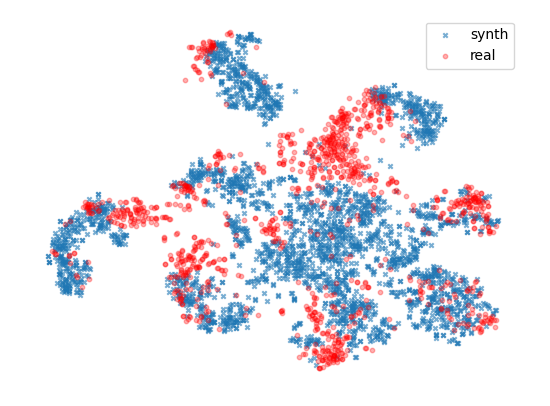}
}
\caption{t-SNE visualization of embedded features (SCT)}
\label{fig:figure5}
\end{figure}

\begin{figure}[H]
\centering
\subfigure[FP-CRNN]{
    \includegraphics[width=5cm]{visualize/fpn.png}
}\subfigure[FP-CRNN + SCMT]{
    \includegraphics[width=5cm]{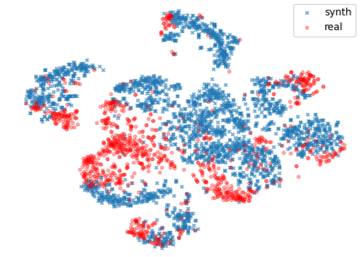}
}\subfigure[FP-CRNN + SCMT + ADA]{
    \includegraphics[width=5cm]{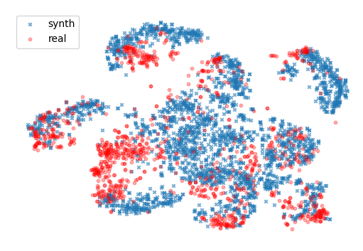}
}
\caption{t-SNE visualization of embedded features (SCMT)}
\label{fig:figure6}
\end{figure}

First, to verify the effectiveness of domain adaptation, we visualize the embedded features before and after domain adaptation using t-SNE. By inspecting the distributions of real and synthetic domain in Figure \ref{fig:figure3}, one may be inclined to conclude that domain adaptation has ability to narrow the domain gap. Further, we conducted the embedded feature analysis on our training strategies using t-SNE individually. Figure \ref{fig:figure4}, \ref{fig:figure5} and \ref{fig:figure6} show embedded features distribution using different training techniques. In Figure \ref{fig:figure4}, We observe that training with ICT tended to widen the gap between synthetic and real data, which is opposite to what domain adaptation did in Figure \ref{fig:figure3}. Moreover, the domain gap can still be clearly recognized after applying ADA. In contrast, though the difference between the distributions of real and synthetic audio data is noticeable when simply using FP, or combining FP with SCT or SCMT, the situation seems not as extreme as the results produced by training with ICT.

To quantitatively justify the observations above, we used silhouette analysis \cite{silhouette} to evaluate the level of dispersion between the feature distributions of real and synthetic data. For each data entry $i$, the silhouette coefficient $s_i$ is defined as follows,
\begin{equation}
    s_i = \frac{b_i - a_i}{\max{(a_i, b_i)}},
\end{equation}
where $a_i$ denotes the average distance between data point $i$ and all other data within the same cluster, and $b_i$ denotes the lowest average distance between data point $i$ and all the points in any other cluster.

The value of $s_i$ ranges from $-1$ to 1. A value close to 1 indicates that the sample is far away from neighboring clusters. If the silhouette coefﬁcient is close to 0, it indicates that the sample is close to the decision boundary between two neighboring clusters. Finally, a negative value indicates that the sample might be in a wrong cluster. 

The average of silhouette coefficient over all samples is called the \emph{silhouette score}. By calculating the silhouette score of the t-SNE feature maps clustered by the domain labels (real or synthetic), we can quantify the extent to which the model pulls the domains apart. The results are shown in Table \ref{tab:table4}; note that
the scores of FP-CRNN with ICT are higher than other SSL strategies combinations, which means it has the tendency to separate real and synthetic domains. Also note that, by applying ADA to the model, the silhouette score can be decreased, which means that ADA successfully encourages the feature extractor to produce domain-invariant embedded features. 


\begin{table}[h]
\caption{F1 (\%) and Silhouette score comparison of different SSL strategies before and after domain adaptation}
\label{tab:table4}
\begin{center}
\begin{tabular}{ccccccc}
\hline 
Architecture & SSL Strategies & ADA  & F1 (\%) & \textbf{Silhouette score}  \\ \hline \hline
FP-CRNN &   -   &   -   &   40.8   &    0.11    \\
FP-CRNN &   -   &   v   &   42.1   &    0.05    \\ \hline
FP-CRNN &   ICT  &   -   &   43.2   &   \textbf{0.23}    \\
FP-CRNN &   ICT  &   v   &   43.9   &   \textbf{0.24}    \\ \hline
FP-CRNN &   SCT  &   -   &   42.7   &   0.09    \\
FP-CRNN &   SCT  &   v   &   43.7   &   0.04  \\ \hline 
FP-CRNN &   SCMT  &   -   &   44.7   &  0.05    \\
FP-CRNN &   SCMT  &   v   &   47.2   &  0.03    \\ \hline

\end{tabular} 
\end{center}
\end{table}

By investigating the implementation of ICT in our previous work, we noticed that Eq.~(\ref{eq:ICT}), which only uses real unlabeled data, boosts the performance by generating and training on ambiguous data in the real domain itself. We suspect that this process concentrates the distribution of the real domain, while pushing the distributions of real and synthetic audio data away from each other.
Consequently, it leads to the consistency within synthetic or real domains alone, but not the consistency in the entire dataset. This being said, however, ICT and ADA has been successfully integrated in the field of computer vision \cite{ADA_ICT}. We believe that the mutual compatibility of ADA and ICT is a topic worth deeper investigation in sound event detection in the future. 
\section{\uppercase{Conclusions}}
In this work, we demonstrate a novel combination of learning strategies for SED by (1) replacing SCT with SCMT and (2) removing the ICT, which was shown to conflict with domain adaptation according to t-SNE visualization and silhouette analysis. Hence, the proposed model surpassed the best-performing model from our previous work by 2.1\% in terms of F1-score on the DCASE 2020 task 4 dataset. 

\section*{\uppercase{Acknowledgements}}
This research is supported by the Ministry of Science and Technology of Taiwan (Grant No.~110-2634-F-002-050).

\renewcommand{\refname}{\normalfont\selectfont\normalsize}
\renewcommand{\refname}{REFERENCES}
\bibliographystyle{vancouver}
\bibliography{main}
\vspace{-18pt}
\end{document}